\documentclass[prl,twocolumn,10pt,aps,superscriptaddress,groupedaddress,longbibliography]{revtex4}  

\usepackage{graphicx}  
\usepackage{dcolumn}   
\usepackage{bm}        
\usepackage{amssymb} 
\usepackage{amsmath}
\usepackage{commath}
\usepackage{mathrsfs}
\usepackage{paralist}
\usepackage{color}
\usepackage{courier}

\begin{document}
\title{Magnetic Levitation Stabilized by Streaming Fluid Flows}
\author{K. A. Baldwin}
\affiliation{Max Planck Institute for Dynamics and Self-Organization, 37077 G\"ottingen, Germany}
\affiliation{School of Science and Technology, Nottingham Trent University, Nottingham, NG11 8NS, UK}
\author{J.-B. de Fouchier}
\affiliation{School of Science and Technology, Nottingham Trent University, Nottingham, NG11 8NS, UK}
\author{P. S. Atkinson}
\affiliation{School of Science and Technology, Nottingham Trent University, Nottingham, NG11 8NS, UK}
\author{R. J. A. Hill}
\affiliation{School of Physics and Astronomy, University of Nottingham, Nottingham, NG7 2RD, UK}
\author{M. R. Swift}
\affiliation{School of Physics and Astronomy, University of Nottingham, Nottingham, NG7 2RD, UK}
\author{D. J. Fairhurst}
\email[]{david.fairhurst@ntu.ac.uk}
\affiliation{School of Science and Technology, Nottingham Trent University, Nottingham, NG11 8NS, UK}
\date{\today}

\begin{abstract}
We demonstrate that the ubiquitous laboratory magnetic stirrer provides a simple passive method of magnetic levitation, in which the so-called `flea' levitates indefinitely.
We study the onset of levitation and quantify the flea's motion (a combination of vertical oscillation, spinning and ``waggling''), finding excellent agreement with a mechanical analytical model. The waggling motion drives recirculating flow, producing a centripetal reaction force that stabilises the flea. 
Our findings have implications for the locomotion of artificial swimmers, for the development of bidirectional microfluidic pumps and provide an alternative to sophisticated commercial levitators.
\end{abstract}
\maketitle

\begin{figure*}[ht]
\includegraphics[width=180mm]{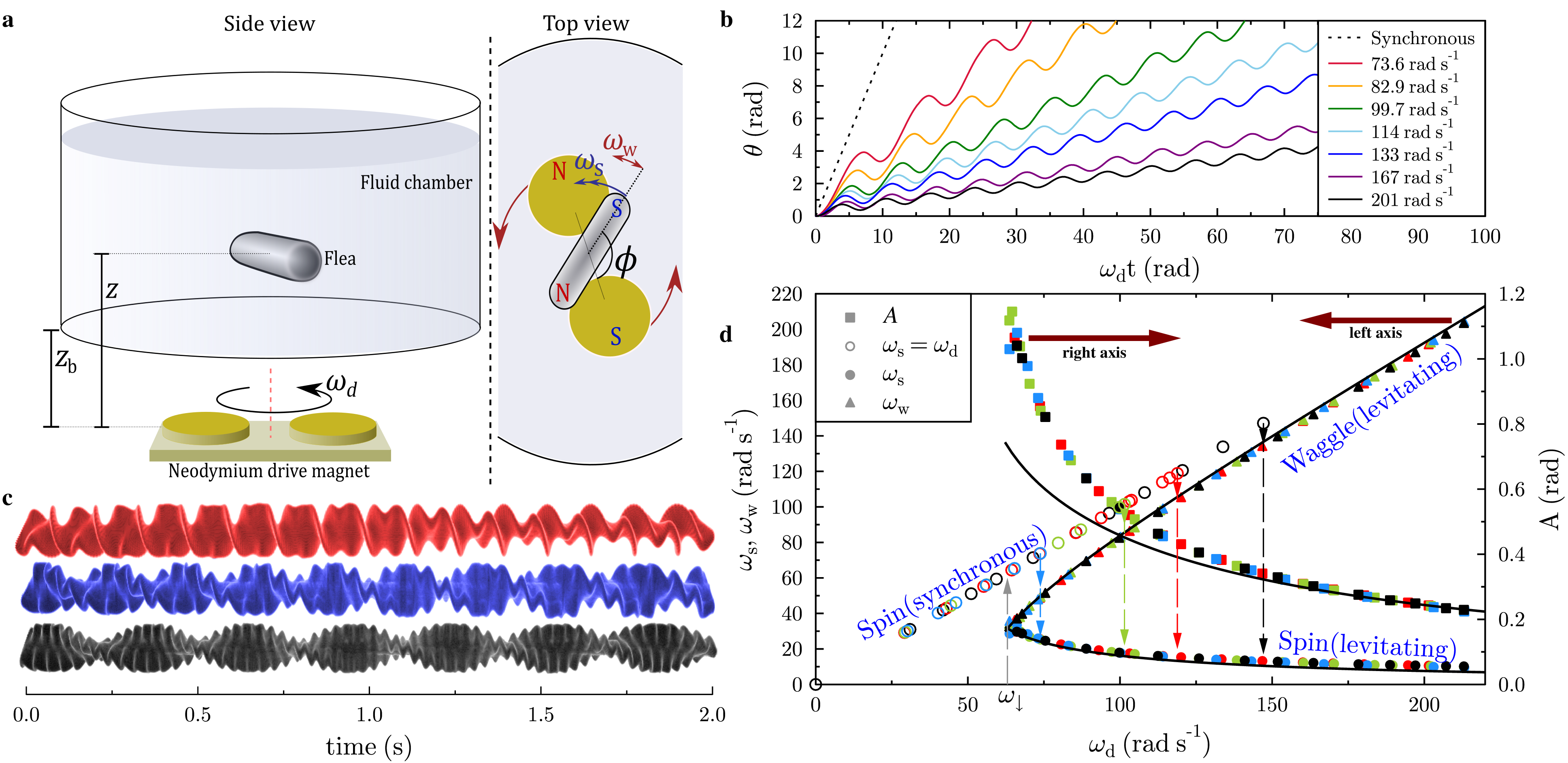}
\caption{\label{3D}{{a}}, Experimental setup showing the flea levitating at $z>z_\text{b}$. 
b, Measured flea angle $\theta$ versus drive magnet angle $\omega_\text{d}t$, for a range of $\omega_\text{d}$. The dotted line corresponds to the synchronously spinning flea when $\omega_\text{d}<\omega_\uparrow$.
c, Overhead 2D images of the flea have been stacked to form 3-dimensional space-time spirals, visualizing the waggling and rotational motion of the flea over a 2~s period, at $\omega_\text{d}=73.6$, 133, and $201~\text{rad}~\text{s}^{-1}$ (from top to bottom). Colors correspond to the relevant curves in panel b.
d, Waggle speed ($\omega_\text{w}$), spin speed ($\omega_\text{s}$), and waggle amplitude ($A$) versus $\omega_\text{d}$. Down dashed arrows indicate the threshold drive speed $\omega_\text{d}=\omega_\uparrow$ is reached, as  $\omega_\text{d}$ is increased from 0, when the flea jumps up from its initial position on the base $z=z_\text{b}$ to a stable levitation point at $z>z_\text{b}$. Colors of data points and arrows correspond to $z_\text{b}=22$ (black), 26 (red), 30 (green), and 34 (blue) mm.
Black lines are analytic solutions using the experimental value $\omega_\downarrow=63\text{~rad~s}^{-1}$.
} 
\end{figure*}

Levitation is the technique of applying magnetic, electric or acoustic fields to suspend an object in stable mechanical equilibrium against gravity.
Finding cheap and simple methods for stable levitation offers prospects for widespread applications, \emph{e.g.}, frictionless transport, containerless storage, contact-free manipulation. For magnetic levitation, one must consider Earnshaw's theorem~\cite{Earnshaw}, which states that dipoles can only be levitated if they are dynamically stabilized. 
This has been achieved using superconductors~\cite{Brandt_SuperconductingLevitators}, in maglev trains~\cite{lee2006MAGLEVreview} (with active feedback), in the levitating spinning top (where gyroscopic forces provide stability~\cite{berry1996levitron}), using high fields generated by powerful electromagnets~\cite{baldwin2015artificial,simon2001diamagnetically,valles1997stable} (including in the infamous frog~\cite{berry1997flying} which levitates due its diamagnetism, \emph{i.e.}, the response of the orbital motion of the electrons to the applied magnetic field), 
or using the magneto-Archimedes effect~\cite{mirica2009measuring,subramaniam2014noncontact}. 
Here, we discuss our discovery of a new route to passive magnetic levitation using a standard laboratory tool: the magnetic stirrer. Using this device, we have observed that a simple bar magnet can undergo a transition from stable spinning to a stable oscillatory levitating mode, the dynamics and stability of which are the focus of this manuscript.

The magnetic stirrer has evolved little since its invention in 1942, consisting, in its simplest form, of two spinning bar magnets, where the dipoles are aligned horizontally, one directly above the other. One is driven by an electric motor (the `drive' magnet), and the second, the stir-bar, is submerged in a fluid. When driven too fast, the stir-bars are known to move asynchronously (`spin-out') and hop erratically -- hence their nickname `flea'.

In our set-up we place a flea centrally on the base of a cylindrical container of a homogeneous fluid, directly above the permanent drive magnet which is spun by an electric motor at speed $\omega_\text{d}$, as shown in Fig.~\ref{3D} a (see~\cite{SM} for more details). When stationary the drive and flea magnets align anti-parallel, with phase angle $\phi=\pi$ between them. 
As the drive speed is increased, the flea spins about an axis perpendicular to its longest axis, synchronously with the drive magnet, at spin speed $\omega_\text{s}=\omega_\text{d}$, but with a reduced phase angle ($\phi<\pi$, as depicted in Fig.~\ref{3D} a) due to the viscous torque acting against its motion; we vary the viscous torque via the drive speed and viscosity of the fluid, and vary the initial dipole-dipole coupling via the height of the base of the container above the drive magnet, $z_\text{b}$ (Fig. \ref{3D} a).
We increase $\omega_\text{d}$ slowly to limit inertial effects from the flea's resistance to angular acceleration.
Above a critical threshold speed, the viscous torque lowers the phase angle below $\pi/2$, whereupon the vertical magnetic force becomes repulsive. 
In this regime, we observe three types of asynchronous motion ($\omega_\text{s}\neq\omega_\text{d}$) depending on the experimental parameters. (I) In low viscosity fluids ({\emph{e.g.}} water) we reproduce the chaotic hopping from which the flea derives its name. (II) For higher viscosity fluids ($\eta\gtrsim0.4~\text{Pa.s}$) and $z_\text{b}$ above a threshold value ($z_\text{b}\gtrsim4~\text{cm}$), the drive magnet periodically overtakes the flea, resulting in flea motion which is a superposition of spinning (at $\omega_\text{s}$) and `waggling' (at $\omega_\text{w}$). 
(III) For $\eta\approx0.4$ Pa.s, and for $z_\text{b}\lesssim4$ cm, the vertical magnetic repulsion overcomes gravity, and the flea jumps up to levitate stably up to several centimetres above the base of the container. 
In this type of motion, as in (II), the flea's angular motion $\theta(t)$ is a combination of spinning and waggling, where the waggle speed increases with $\omega_\text{d}$, while the rotation speed decreases. 
For shallow or low viscosity liquids, magnetic stirrers can induce significant vortex flows~\cite{halasz2007vortex}, but we avoid these situations and see no deformation of the liquid surface. 

Fig.~\ref{3D} b shows a plot of experimentally measured flea angle $\theta$ in the levitating state, for various $\omega_\text{d}$. Increasing $\omega_\text{d}$ increases the waggle speed $\omega_\text{w}$ and decreases the spin speed $\omega_\text{s}$. 
Fig. 1 c shows 3D surfaces created by combining images of the flea (viewed from above) over a 2 second period.
The flea's asynchronous motion, for all $\omega_\text{d}$, is well fitted by the empirical equation
\begin{equation}
\label{empiricalequation}
\theta = \omega_\text{s}t +A \sin(\omega_\text{w}t),
\end{equation}
where $A$ is the amplitude of the waggle. The fitting parameters give experimentally obtained values for $\omega_\text{s}$, $\omega_\text{w}$ and $A$ as a function of $\omega_\text{d}$, which are plotted as data points on Fig.~\ref{3D} d.
Once levitating, the angular motion of the flea is independent of initial vertical position (\emph{i.e.} the height of the base $z_\text{b}$), as shown by the collapse of the data in Fig.~\ref{3D} d, implying negligible wall effects from the base. 
When reducing $\omega_\text{d}$ while the flea is levitating, it becomes unstable and falls to the base at $\omega_{\text{d}}=\omega_\downarrow\approx63~\text{rad}~\text{s}^{-1}$ (Fig.~\ref{3D} d), when $\omega_\text{s}=\omega_\text{w}$.

To capture the essential features of the flea's angular dynamics, we model it as a cylinder oscillating about an axis passing through its geometric centre and perpendicular to its long axis, and coinciding with the rotation axis of the drive magnet.
Under this assumption we propose the following equation for the angular motion, which combines the flea's inertia, the viscous torque and the magnetic coupling:
\begin{equation}
\label{thetaequation}
{I}\ddot{\theta}+{D}\dot\theta-{M(z)}\sin(\theta-\omega_\text{d} t)=0,
\end{equation}
where $I$ is the moment of inertia of the flea. 
$D$ is the drag constant for a prolate ellipsoid (approximating that of a cylinder), given by $D=8\pi\gamma K\eta l^3$, where $K$ is a geometric factor (Eq. (18) of ref.~\cite{kong2014swimming}) equal to 0.212 for our flea, $\gamma$ accounts for the increase of drag due to the proximity of the base of the container and $l=12$~mm is half the length of the flea. Our supplemental experiments show drag is proportional to $\dot\theta$ at angular speeds relevant to our experiments~\cite{SM}. Assuming point dipoles, the magnetic coupling is $M(z)=\mu_0m_\text{d} m_\text{f}/4\pi z^3$, where $m_\text{d}$ and $m_\text{f}$ are the magnetic moments of the drive and flea respectively, and $\mu_0$ is the magnetic constant. The constants $m_\text{d}$, $m_\text{f}$, $I$, and $\gamma$ were measured experimentally (see~\cite{SM} for more details).

We first consider the solutions to Eq.~(\ref{thetaequation}) for a constant value of $z$, the mean height of the flea. In general $z(t)$ is oscillatory, so that the angular motion is coupled to the vertical motion. Nevertheless considering the angular motion at fixed $z$ gives us some initial key insights.
For synchronous motion, Eq.~(\ref{empiricalequation}) is a trivial solution to Eq.~(\ref{thetaequation}), where $A=0$ and $\omega_\text{s}=\omega_\text{d}$, leading to a relationship between the phase lag and the drive speed, $\sin(\phi) =\omega_\text{d}/\omega_\uparrow$, where $\phi=\theta(t)-\omega_\text{d}t$. Here $\omega_\uparrow = M(z_\text{b})/D$ is the threshold speed for transition to asynchronous motion when the flea is on the base. We measured $\omega_\uparrow$, varying $z_\text{b}$ and viscosity in 15 different experimental configurations, and found that $\omega_\uparrow = (1.14\pm0.04) M(z_\text{b})/D$, in reasonable agreement with the model. This threshold is identical to the synchronous-asynchronous spinning threshold in magnetic nanorod microrheology~\cite{frenkel1955kinetic,korneva2005carbon, 
tokarev2013probing}. 

For asynchronous motion, there are no simple analytical solutions to Eq.~(\ref{thetaequation})~\cite{Abel}. Numerical solutions, however, show that Eq.~(\ref{empiricalequation}) is an approximate solution under steady state conditions.
We now deduce 3 simultaneous equations for the parameters $A$, $\omega_\text{s}$, and $\omega_\text{w}$ in Eq.~(\ref{empiricalequation}) as functions of $\omega_\text{d}$ and $\omega_\downarrow$. 
Firstly, we note that for fixed $z$, Eq.~(\ref{thetaequation}) maps to that of a damped pendulum driven by constant torque (solved for the zero inertia limit by Coullet \emph{et. al}~\cite{coullet2005damped}). Combining their result (Eq.~(14) of ref.~\cite{coullet2005damped}) with our observation that $\omega_\text{d}=\omega_\downarrow$ at $\omega_\text{s}=\omega_\text{w}$, leads to
$\omega_\text{w}^2=\omega_\text{d}^2-\frac{3}{4}\omega_\downarrow^2$. 
Secondly, we determine two expressions for the maximum speed from differentiating Eqns.~(\ref{empiricalequation}) and (\ref{thetaequation}), which when equated give $\omega_\text{s}+A\omega_\text{w}=\frac{\sqrt{3}}{2}\omega_\downarrow$. 
Finally, we note that $\omega_\text{d}=\omega_\text{w}+\omega_\text{s}$. This is because, over the time interval between consecutive waggles given by $\Delta t=2\pi/\omega_\text{w}$, the flea moves $\Delta\theta_\text{f}=2\pi\omega_\text{s}/\omega_\text{w}$, whereas the drive has moved by $\Delta\theta_\text{d}=2\pi\omega_\text{d}/\omega_\text{w}$. $\Delta\theta_\text{d}$ must also be equal to $\Delta\theta_\text{f}+2\pi$, as the phase angle between the flea and the drive must start and finish at the same value over this period, and the flea waggles every time it is lapped by the drive. 

Solving the three simultaneous equations and using the experimental value of $\omega_\downarrow=63 \text{~rad s}^{-1}$, we calculate analytical values, plotted as solid lines on  Fig.~\ref{3D} {{d}}, with no free fitting parameters.
The angular speed data are fitted well by the analytical curves. 
The amplitude is fitted well except at low $\omega_\text{d}$, possibly due to the simplified drag model. 

\begin{figure}
\includegraphics[width=95mm]{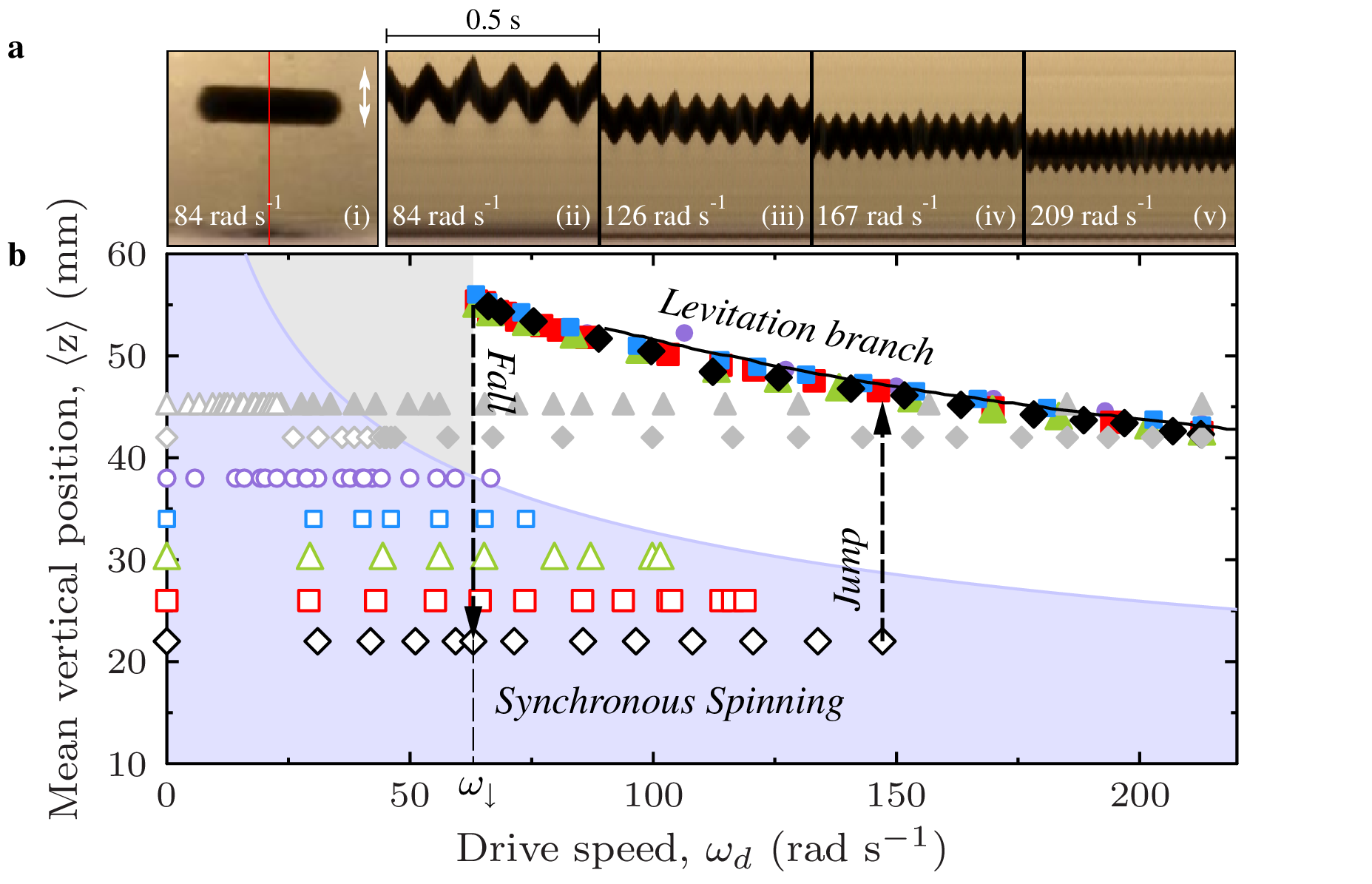}
\caption{\label{Orthogonals}{{a}}, Flea levitating in castor oil (drive speed $\omega_\text{d}$ labelled). (i) a still photograph of the levitating flea, (ii-v) projections of the central pixel column (red line in (i)) over 0.5 s. 
b, Mean vertical position $\langle z\rangle$ of flea versus $\omega_\text{d}$ for range of $z_\text{b}$.
Dashed down arrow shows where $\omega_\text{d}=\omega_\downarrow$ for all experiments. Solid black line: mean vertical position calculated by numerical integration of Eqns.~(\ref{thetaequation}) and~(\ref{zequation}). Plotted symbols show experimental results: (hollow) synchronous spinning; (coloured solid) asynchronous, levitating; (grey solid) asynchronous, non-levitating. Blue shaded region: synchronous spinning. Boundary calculated analytically using $\omega_\uparrow=M(z_\text{b})/D$. Grey shaded region: asynchronous, non-levitating. 
}
\end{figure}

Fig.~\ref{Orthogonals} a shows the vertical motion of the flea. Fig.~\ref{Orthogonals} b shows the experimentally-obtained mean height $\langle z\rangle$ of a levitating flea, which decreases with increasing $\omega_\text{d}$ (also apparent in Fig.~\ref{Orthogonals} a). Also shown are analytically-determined bounds on $\omega_\text{d}$ and $z$ for synchronous and asynchronous motion and comparison with experimental data. Between $\omega_\uparrow$ and $\omega_\downarrow$ the system shows hysteresis depending on how it was prepared: the flea can either be spinning synchronously or levitating (asynchronous angular motion). Consider the hollow black symbols on Fig.~\ref{Orthogonals} b: on increasing the drive speed from stationary, the flea spins synchronously on the base (at $z_\text{b}=22$~mm) until $\omega_\text{d}$ reaches 146 rad s$^{-1}$, whereupon the flea jumps up abruptly from the base to levitate at $\langle z\rangle=46$~mm. On reducing $\omega_\text{d}$ from this point, $\langle z\rangle$ follows the levitation curve shown (solid black symbols), increasing until $\omega_\text{d}<\omega_\downarrow$, whereupon the flea falls and reverts to synchronous spinning on the base. The levitation height $\langle z\rangle$ is not influenced by the proximity of the base, (\emph{i.e.} independent of $z_\text{b}$), except when $z_\text{b}$ exceeds a critical height (approx. 40~mm) such that $\omega_\uparrow<\omega_\downarrow$ (grey symbols); then stable levitation is not possible and the flea waggles on the base in asynchronous motion (case II described above).

Following our consideration of the angular dynamics at fixed height, we now introduce the coupled equation for the vertical motion. We propose the following model for the vertical forces, again assuming coupling between point dipoles (which has a $\sim z^{-4}$ dependency), 
\begin{equation}
\label{zequation}
\frac{\ddot{z}}{g'} + \frac{\dot{z}}{v_\text{t}} - \left(\frac{z_0}{z}\right)^4 \cos(\theta-\omega_\text{d}t) + 1 = 0,
\end{equation}
where $v_\text{t}$ is the flea's translational terminal velocity in the absence of any magnetic forces, $g'$ is the buoyancy-corrected gravitational acceleration and $z=z_0$ is the theoretical equilibrium vertical separation when the two magnets are aligned ($\phi=0$) and stationary ($\omega_\text{d} = 0$). 
Here we have scaled each term by the buoyant weight of the flea.
We use numerical methods to calculate the mean vertical position $\langle z\rangle$ predicted by the coupled $\theta$ and $z$ equations of motion, plotting the solution as a black line with no adjustable parameters on Fig.\ref{Orthogonals} b; $g'$, $v_\text{t}$ and $z_0$ were obtained experimentally. 
We find good agreement between numerical results and experimental data for the mean levitation height, with the flea losing vertical stability at $\omega_\text{d}<\omega_\downarrow$ in both model and experiment.
Levitation requires that the time-averaged vertical magnetic force balances the gravitational force, which occurs in asynchronous motion. 
Experiment and modelling shows that if the drive speed is too slow, the flea's motion synchronises with the drive magnet and the flea falls.
The numerical results predict that the low end of the stable levitation branch ends at $\omega_\downarrow\approx90$~rad~s$^{-1}$, higher than the observed value, possibly caused by overly simplifying the fluid in the container, considering only Stokes viscous drag and ignoring fluid inertia.

\begin{figure*}[ht]
\includegraphics[width=180mm]{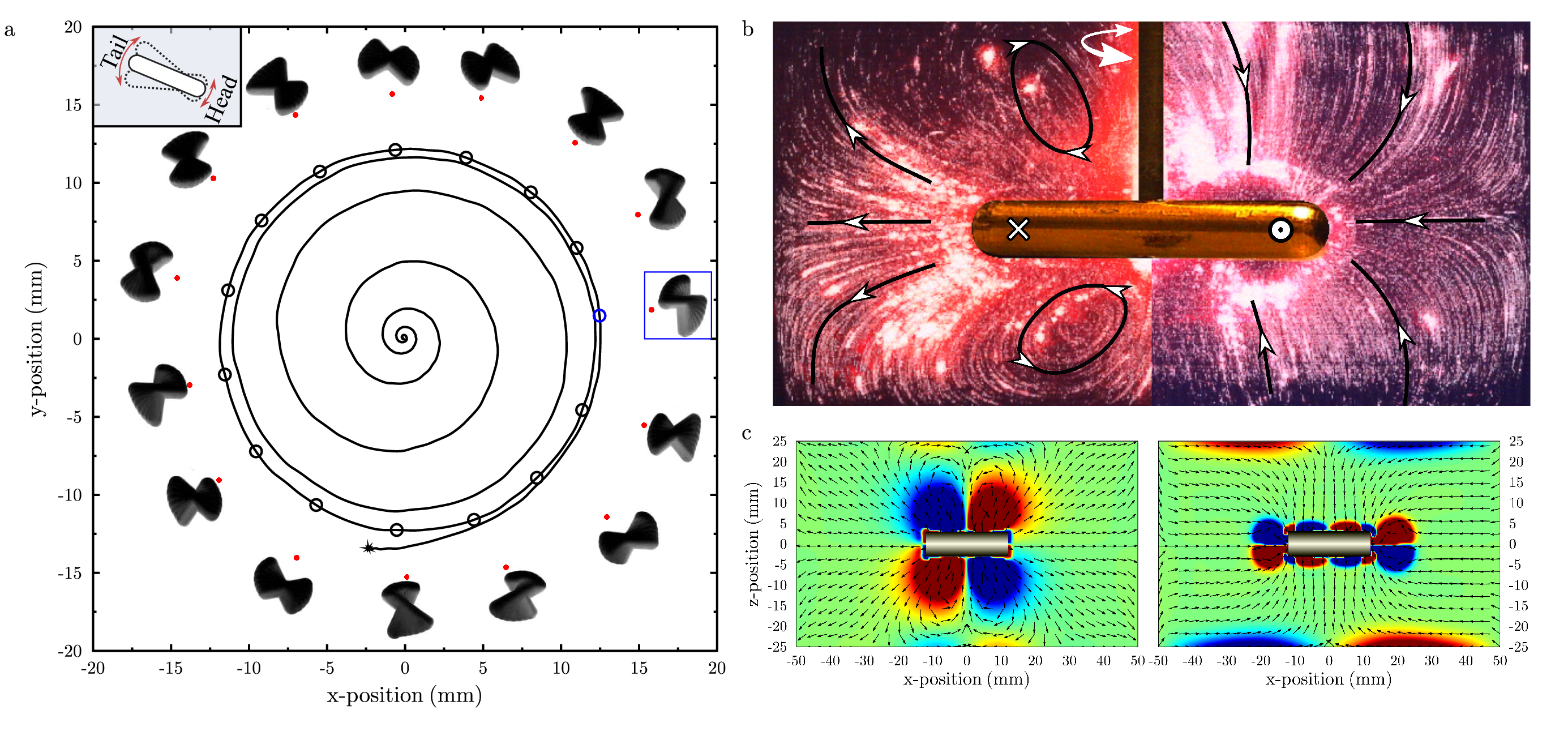}
\caption{\label{Trap}{{a}}, Solid line: experimentally determined path of the flea following displacement of the drive magnet ($\omega_\text{d}=73~\text{rad~s}^{-1}$) 14~mm from the origin (averaged over 7 frames).
The images are projections of the waggle motion as the flea spirals towards the rotation axis of the drive magnet, taken at 15 points, at intervals of 0.14~s (34 frames), corresponding to the positions indicated by circles on the spiral. Red dots indicate the axis of the drive magnet relative to the flea. Inset diagram indicates the `head' and `tail' of the asymmetrically waggling flea when the rotation axes of the two magnets are displaced horizontally. {{b}}, 1~s projections of suspended particle paths (directionality given by black lines with white arrowheads), driven by our ``artificial waggler'' (the ends of the rod are oscillating in and out of the page).
Left and right sides of the image are taken from separate experiments at $\text{Re}_\text{s}=11.7\pm0.4$ (left) and $\text{Re}_\text{s}=400\pm12$ (right). 
Image of brass bar used in experiment has been superimposed for clarity.
c, Plots of the simulated fluid flows: left and right plots represent same $\text{Re}_\text{s}$ as in experimental flows shown in b. Colour map and arrows indicate the vorticity and direction of flows respectively.} 
\end{figure*}

In Eqns.~(\ref{thetaequation}) and (\ref{zequation}), we implicitly constrain the motion of the flea to the same axis as the drive magnet, but in the experiments there is no such constraint. This raises the question: what provides the radial stability? 
In experiments, we observe that the flea is unable to stay centred above the drive magnet below a critical viscosity. Computationally we observe that a simple numerical model of an unconstrained flea that excludes fluid inertia is also radially unstable. Both these observations suggest a complex hydrodynamic origin to the radial stability.

To investigate the radial stability experimentally, we rapidly displace the drive magnet by 14 mm horizontally during levitation, and observe the flea returning to the axis of the drive magnet along a spiral path (Fig.~\ref{Trap} {{a}}) with a mean radial speed of $\approx5~\text{mm s}^{-1}$. During this spiral path, the waggle is eccentric: the end of the flea furthest from the drive rotation axis (`the tail') sweeps through a greater arc in the fluid than the other end (the `head') and drives greater fluid flows (see inset Fig.~\ref{Trap} a).

To determine the directionality of these waggle-induced radial flows and elucidate their effect on stability, we built an `artificial waggler' to reproduce the waggling motion of the flea without the vertical or rotational motion. 
The waggler consists of a motorized rod 
that reproduces the waggle motion of the levitating flea without the slow spin or vertical motion. 
The device allows for precise control of waggle speed and amplitude in the ranges $20<\omega_\text{w}<130$ rad s$^{-1}$ and $\pi/40<A<\pi/4$ rad. We adjust these parameters and the viscosity to control the streaming Reynolds number, which characterizes the ratio of inertial to viscous forces under oscillation-induced streaming flows, given by $\text{Re}_\text{s}=2A^2l^2\rho\omega_\text{w}/\eta$, where $\rho$ is the liquid density.
We imaged the flows (Fig.~\ref{Trap} {{b}}) via pathlines of suspended mica particles illuminated by a collimated laser sheet, at $\text{Re}_\text{s}$ values identical to those in two noteworthy cases: (i) a stably levitating flea at high viscosity ($\text{Re}_\text{s}=11.7\pm0.4$); (ii) an initially levitating flea at low viscosity ($\text{Re}_\text{s}=400\pm12$) which becomes radially unstable, drifting sideways away from the drive's rotation axis, and falling after approximately 30 s. 
We find that there is a striking difference between the flows in the two cases: in (i), the fluid is drawn inwards from both above and the sides of the flea, and pumped outwards along its axis; in (ii), fluid is drawn inwards from both along its axis and above, and pumped outwards to the sides.

Fig.~\ref{Trap} {{c}} shows the flows generated by computational simulations of a model waggler, with dimensions and fluid parameters the same as those in the experiments (but within a smaller container due to numerical limitations). These simulations are based on the embedded boundary method described previously~\cite{klotsa2007interaction,pacheco2013spontaneous,klotsa2015propulsion} (here we used 0.25 mm lattice spacing and 0.01 ms time-step - see~\cite{SM} for more details), which show good qualitative agreement with experiment. Additionally, we used these simulations to calculate the net resultant force, time averaged over one cycle, acting on a flea under the same conditions, but driven to oscillate eccentrically. 

At $\text{Re}_\text{s}=11.7$ (stable levitation, outward flow), the force acting on the flea quickly reaches a steady value of -0.58 mN. Here, a negative value indicates that the force acts to propel the flea in the direction pointing from the `tail' to the `head'.
This would be stabilizing for the circling flea shown in Fig.~\ref{Trap} {{a}}, propelling it toward the drive's rotation axis.
On the other hand, for $\text{Re}_\text{s}=400$ (unstable levitation, inward flow), we find that while the resultant force is initially negative, as the fluid flow settles over a period of around 15 s the force transitions to a steady positive value of $0.26~\text{mN}$, \emph{i.e.}, a destabilizing force (see~\cite{SM} Fig. S1).
Similar flow reversal has been observed around an oscillating sphere due to a change in the thickness of the oscillatory bounding layer~\cite{riley1966sphere,otto2008measurements} consistent with our simulations.

In summary, we have discovered a new route to stable levitation using an inexpensive and readily-available laboratory tool: the magnetic stirrer. We demonstrate experimentally and in simulations that above a critical drive speed, the flea's angular motion desynchronises from the drive resulting in a net vertical magnetic force that levitates the flea. Our experiments and simulations lead us to propose that levitation is stabilised by an asymmetric fluid flow, driven by the flea's eccentric swim stroke when the flea moves off axis. This only occurs at intermediate streaming Reynolds numbers where the flow is pumped radially outwards; at higher streaming Reynolds number, the flow reverses, and levitation is unstable.
We anticipate that this flow-switching at intermediate streaming Reynolds numbers will have prospects for the design of novel bi-directional fluidic pumps, and for understanding artificial swimmers~\cite{williams2014self,palagi2016structured,wang2012inertial} in this relatively poorly understood intermediary fluid regime. Further, this novel combination of levitation plus induced fluid flow could lead to new approaches for homogenous surface treatment, or dynamic viscosity measurements.

D.J.F. acknowledges the Erasmus scheme for funding J.-B.d F., and Nottingham Trent University for funding both a year-long sabbatical and personnel funding for P.A. and K.A.B.~~R.J.A.H. acknowledges support from an EPSRC Fellowship, Grant No. EP/I004599/1. The authors collectively thank Dave Parker and Dave Holt for building custom experimental apparatus, and Stephen Wilson and Brian Duffy for highlighting the connection with Abel equations.


\begin{thebibliography}{33}
\bibitem{Abel}
This equation can be mapped onto an Abel equation of the second kind, but with no known solution.
\end{thebibliography}

\end{document}